# The Golden Ratio as a proposed solution of the Ultimatum Game: An explanation by continued fractions

Stefan Schuster

Friedrich Schiller University Jena

Dept. of Bioinformatics

Ernst-Abbe-Platz 2

07743 Jena, Germany

E-mail: stefan.schu@uni-jena.de

**Abstract**

The Ultimatum Game is a famous sequential, two-player game intensely studied in Game Theory. A proposer can offer a certain fraction of some amount of a valuable good, for example, money. A responder can either accept, in which case the money is shared accordingly, or reject the offer, in which case the two players receive nothing. While most authors suggest that the fairest split of 50% vs. 50% would be the equilibrium solution, recently R. Suleiman (An aspirations-homeostasis theory of interactive decisions (2014) http://vixra.org/abs/1403.0029) suggested the Golden Ratio, 0.618…, to be the solution and argued that such a partitioning would be considered fair by both sides. He provided a justification in terms of an approach termed aspirations-homeostasis theory. The main idea is that responders tend to accept the minor fraction of the Golden Ratio because they feel that this fraction equals, in comparison to the larger fraction obtained by the proposer, the ratio of the larger fraction and the whole amount. Here we give an alternative explanation to that suggested solution, which complements the reasoning by Suleiman (2014) and is based on infinite continued fractions.

**Introduction**

The Ultimatum Game is a famous asymmetric, two-player game intensely studied in Game Theory. It had been devised by Güth et al. (1982). The idea is the following: One player, called the "proposer", is handed a valuable good, say 100 € She is to offer any part of it to the second player, called the "responder". We here use the feminine forms (she, her etc.) throughout, meaning to cover both genders. The responder can choose between two strategies: to accept or to reject. If she accepts, the money is shared accordingly. If she rejects, neither of the players receives anything. It is a sequential game because one player makes her move first and the other one only thereafter. There are several modified versions such as the δ-Ultimatum Game (Suleiman, 1996) and a spatial variant (cf. Iranzo et al., 2011).

In the standard Ultimatum Game, there is a case that at very low total amounts, say 10 cents, any non-zero offer would be accepted because the amount is so tiny. Also at very large total amounts, say 10 million €, low fractional offers, say 10 % might be accepted because the responder might not dare to decline 1 million €. However, experimental observations show that the outcome changes only insignificantly if stakes are raised (cf. Camerer and Thaler, 1995). Thus, in what follows, we assume that the total amount does not really matter and we argue in terms of percentages.

We denote the fraction the proposer wants to keep by $x$. Then the offer is $1-x$. If we assume that the smallest unit of the used currency is such that the smallest non-zero offer is 1%, then $x = 99\%$, $1-x = 1\%$ is the Nash equilibrium, the so-called rational solution (cf. Page and Nowak, 2000, 2002). This is because the responder does not have an incentive to decline that offer, as it is still larger than 0. The proposer would not deviate either from that strategy because it grants her the largest amount. However, the manifold experiments with that game in diverse ethnical communities show that such a low offer is usually rejected. Roth et al. (1991) run the Ultimatum Game in four different cities: Jerusalem, Ljubljana (Slovenia), Tokyo and Pittsburgh. The results were strikingly similar. In all four locations, the modal offers were in the range of 0.4–0.5 and were usually accepted by the responders. Henrich et al. (2005) conducted a very comprehensive study with 17 different ethnic groups all over the world. The average offers ranged from 0.26-0.57 with a pronounced peak in the range 0.4-0.45 (9 ethnic groups).

Offers below 0.3 (cf. Nowak et al., 2000) or below 0.2 are usually rejected (Page and Nowak, 2002; Suleiman et al., 2014). A clear-cut threshold is hard to determine because there is actually a continuous frequency distribution of rejection as a function of $x$.

The Ultimatum Game was also played with chimpanzees, using raisins as the good to be shared (Jensen et al., 2007; Proctor et al., 2013). It turned out that the Nash equilibrium (rational solution) describes their behaviour quite well, that is, responders were satisfied with low offers (for a critical discussion, see Henrich and Silk, 2013).

Although the Ultimatum Game is a one-shot game, most people assume that it will be iterated (cf. Nowak et al., 2000). An alternative reasoning leading to the same conclusion is that people have learned rules of behaviour from iterated games that they apply also to one-shot games (Camerer and Thaler, 1995). This may explain why very low offers are rejected. Responders tend to "punish" proposers who make low offers, hoping that they will make larger offers in the next round. Note that the proposer in the next round need not be the same person. Since, usually, people in a population can communicate, all people know that selfishness would be punished and would imply a low reputation (cf. Nowak et al., 2000).

Some authors distinguish between two parameters: $p$ as the average offer (denoted here by $1-x$) and $q$ as the minimum offer responders will accept. Computer simulations by Page and Nowak (2000) show that $q$ tends towards $p$ in evolutionary dynamics. This is intuitively understandable because if $q < p$, proposers would decrease $p$. Accordingly, we need only one parameter to describe the solution, at least approximately. A quantitatively more precise approach would be to derive frequency distributions describing the acceptance/rejection behaviour of the proposer and responder, respectively.

Game Theory has become a large and powerful theoretical framework. Entities ranging from nations, companies (cf. Gintis, 2000), staff headquarters, people (cf. Güth et al., 1982; Page and Nowak, 2000, 2002), animals (Maynard Smith, 1982; Hofbauer and Sigmund, 1998), plants (King, 1990) and living cells (Pfeiffer and Bonhoeffer, 2004) down to molecules (cf. Bohl et al., 2014) are considered as players. The tools used in that theory are quite diverse as well. Not only the classical Nash equilibrium determined in payoff matrices is used, but also differential equations (Hofbauer and Sigmund, 1998; Gintis, 2000), the concept of evolutionary stable strategies (Maynard Smith, 1982; King, 1990) and others.

Recently, Suleiman (2014) suggested that the Golden Ratio (GR) would be a plausible solution of the Ultimatum Game. This is intuitively appealing because the Golden Ratio, 0.618…, is one of the few numbers between 0.5 and 1 that can be explained in a minimalist, natural way. One out of several definitions for that number is the equation $(1-x)/x = x$. It says that the ratio between the smaller and larger fractions is the same as the ratio between the larger fraction and the whole. This leads to the solution

$$x^* = \frac{\sqrt{5}-1}{2} \cong 0.618...\qquad(1)$$

As $\sqrt{5}$ is an irrational number, so is the GR. The GR is used frequently in architecture, painting and photography and occurs in phyllotaxis (leaf positions on plants) (Jean, 1994).

Another definition is based on the Fibonacci numbers. These numbers are defined by the recursive equation

$$f_{n+1} = f_n + f_{n-1}\qquad(2)$$

and the initial conditions

$$f_1 = 1, \ f_2 = 1 \ . \tag{3}$$

This leads to the series 1, 1, 2, 3, 5, 8, 13, … It can easily be shown that the ratio of two consecutive Fibonacci numbers tends to the GR (cf. Koshy, 2001):

$$\lim_{n \to \infty} \frac{f_{n-1}}{f_n} = x^* \ . \tag{4}$$

The main idea of Suleiman (2014) in his justification for the proposed solution is based on definition (1). Responders tend to accept an offer of the minor fraction of the GR, $1-x^*$, because they feel that this fraction corresponds, in comparison to the larger fraction obtained by the proposer, to the ratio of the larger fraction and the whole amount; the latter is the maximum amount the proposed could get in principle. This equality of fractions may be felt by both players as a fair division. In an evolutionary context, this might be related to the observation that relative fitness (that is, in comparison to competitors) is more important than absolute fitness.

Earlier authors (e.g. Nowak et al., 2000; Page and Nowak, 2001, 2002), in contrast, call the 50 % : 50 % solution the fair split. To distinguish the two, we will call the 50 % : 50 % solution "equipartition".

Here, we will give another explanation for the above-mentioned hypothesis that the GR would be the equilibrium solution of the Ultimatum Game. It is based on continued fractions, a tool used in several areas in mathematics (Olds et al., 1992; Cuyt et al., 2008).

**Explanation in terms of continued fractions**

When a proposer can make an offer to the responder, she will certainly think about two possible options:

(i) How would the responder react if I offered $x_1 = 1$? Answer: She would certainly decline.

(ii) How would the responder react if I offered $x_2 = 0.5$? Answer: She would certainly accept because this is the fairest division.

These two thought experiments need not be performed in the given order. This does not, however, affect our line of reasoning.

The offers mentioned in (i) and (ii) can be written as

$$x_1 = \frac{1}{1}, \quad x_2 = \frac{1}{1+1} \tag{5a,b}$$

respectively.

Although the division of money mentioned in (ii) is the fairest deal at first sight, there is a case for assuming that it is not the stable solution. Consider a population of responders and assume first that all of them would decline any offer below 0.5. This would force proposers to stay with equipartition. However, a subpopulation of responders could start accepting offers lower than 0.5. This strategy would grant them an advantage, at least short-term. Under the plausible assumption that there are always some fluctuations in the offers (cf. Nowak et al., 2000), proposers will notice that there are some responders accepting offers lower than 0.5, and therefore, will make them. As those responders who decline such offers will not get anything, the modest responders will have an advantage. As they undermine the punishing strategy, they can, in a sense, be considered as cheaters. However, we should be careful with such moral notions, all the more as we could, paradoxically, call these responders both modest (or frugal) and selfish. Finally, all responders must act in this way because the cooperative responders (accepting only offers $x \geq 0.5$) are outcompeted. Thus, the evolutionarily stable strategy (ESS, Maynard Smith, 1982) is certainly to accept offers lower than 0.5.

This situation is reminiscent of the arms race in the growth of trees in a forest. It would be best for all trees in a forest to reach exactly the height optimal for biomass production. However, this is not an ESS. It can be invaded by trees that are taller because they get more sunlight. The evolutionarily stable height is somewhat larger than the optimum height (King, 1990).

The question arises down to what minimum value offers can decrease so that they are still accepted by the modest responders. A straightforward question proposers will ask themselves is:

(iii) How would the responder react if I offered him half of what I get? That is, $x_3 = 2/3$.

The rationale for (iii) is that the ratio $(1-x)/x = ½$ would be that positive ratio less than unity that involves the smallest integers possible.

The answer by the responder is more difficult to predict. There is a case for her declining because she may be disappointed to get only half of what the proposer gets. Empirical observations show that some responders accept but many decline. Anyway, the fraction $x$ can, in this case, be written as

$$x_3 = \frac{2}{3} = \frac{1}{1+\dfrac{1}{1+1}} \tag{6}$$

Inspecting the formulas (5a,b) and (6) and extending the reasoning in a straightforward way, one may conclude that the equilibrium solution is given by the infinite continued fraction

$$x^* = \frac{1}{1+\dfrac{1}{1+\dfrac{1}{1+\dfrac{1}{...}}}} \cong 0.618 \tag{7}$$

This result can be substantiated in a more detailed way as follows. The proposer could ask herself whether a small deviation from $x_2 = 0.5$ would be acceptable by the responder, say $x = 0.51$. It probably would because the offer is very near to 0.5 and the responder is likely to regret renouncing to take 0.49 just because of the small deviation. In other words, she will approximate 0.49 by 0.5. Now, the same reasoning can be applied to $x = 0.52$, $x = 0.53$ etc. The reasoning applies as long as $x$ can still be considered as an approximation of 0.5. In contrast, $x = 0.65$, for example, could not be considered that way anymore, because it is much nearer to $x_3 = 2/3$. The equilibrium solution will be that number that is in between 0.5 and 2/3 and is hardest to approximate by these values. This leads to the GR. That number is even hardest to approximate by any rational numbers in the sense that it is furthest away from any ratio of two small integers. On the basis of that definition of "irrationality", it can in fact be proved that the GR is the most irrational of all numbers (cf. Havil, 2012).

That this infinite fraction really gives the GR can be seen as follows. As any infinite part of the fraction is the same as the entire fraction, we can write

$$x^* = \frac{1}{1+\dfrac{1}{1+\dfrac{1}{1+\dfrac{1}{...}}}} = \frac{1}{1+x^*} \tag{8}$$

This implies $x*^2 + x* - 1 = 0$, which gives the solution (1). Strictly speaking, it should be proved, in addition, that the continued fraction really converges. This can easily be seen by realizing that the values $x_1$, $x_2$, and $x_3$ given above in the form of fractions are steps towards the infinite fraction. Moreover, they are quotients of Fibonacci numbers. In fact, all the finite fractions involved in the infinite fraction (6) are quotients of Fibonacci numbers (cf. Olds et al., 1992; Cuyt et al., 2008; Havil, 2012). For example,

$$x = \cfrac{1}{1 + \cfrac{1}{1 + \cfrac{1}{1+1}}} = \frac{3}{5} \; . \tag{9}$$

The above-mentioned criterion of being hard to approximate by rational numbers exactly corresponds to the infinite continued fraction (7). It is related to the observation that the infinite fraction converges slowly because it only contains the lowest positive integer, unity, so that the partial denominators are small (cf. Cuyt et al., 2008).

**The 60:40 approximate solution**

As the proposer has to express her offer in usual numbers rather than by mentioning the GR or the square root of 5, she will usually round the offer. Due to our decimal numeral system, a straightforward way of rounding the GR is 60 %, thus offering 40 % to the responder. As 60 % = 3/5, this is again equivalent to a ratio of Fibonacci numbers, as can be seen in Eq. (9). This coincidental equality of the rounded value with a ratio of Fibonacci numbers makes this approximate solution very appealing. It is in agreement with a frequently observed offer (cf. Henrich et al., 2005; Suleiman, 2014).

It is worth noting that there are other number series that lead to the GR as well. For example, the Lucas numbers 1, 3, 4, 7, 11, … (cf. Koshy, 2001) are defined by the recursion formula (2) as well, but start with

$$f_1 = 1, \; f_2 = 3 \; . \tag{10}$$

The ratio of two consecutive Lucas numbers tends to the GR as well (cf. Koshy, 2001). For example, 7/11 = 0.6363… Let us check whether such ratios could provide a convenient approximate solution to the Ultimatum Game. $x = 1/3$ would certainly be too small, $x = 3/4$ would probably be too high to provide a general solution, although an offer of 25 % is sometimes made. 4/7 = 0.571… and 7/11 = 0.636… are candidate solutions but are much less convenient than 3/5 because the latter matches our decimal numeral system much better.

**Discussion**

Here we have given an explanation to the hypothesis that the Golden Section (GR) could be the solution of the Ultimatum Game. It is appealing due to its good agreement of the theoretical result with experimental observations and the logical flow of reasoning. It is alternative and complementary to the explanation given by Suleiman (2014). That the GR is an irrational number should not be problem. When, for example, a painting or photograph involving the GR is aesthetically appealing, this is realized by people by intuition rather than iteration. This shows that the human brain is capable of estimating such numbers excellently without using a calculator.

Comparing the value of the lower fraction of the GR, 0.382, with the empirical values from the literature (cf. Introduction) shows a remarkable agreement. Inspecting the observations reported by Henrich et al. (2005) in more detail shows that the mean value of offers observed with the ethnical group of Gnau in Papa New Guinea is closest to that value, because it was 0.38. Also very close are the average offers observed with the Tzimane (0.37), Kazakh (0.36) and Hadza (0.40). An observation reported by many authors is that offers below 0.3 are generally rejected (cf. Nowak et al., 2000). This can be explained better by the GR solution than by a 2/3 vs. 1/3 solution or an equipartition. In numerical computer simulations by Nowak et al. (2000), the threshold of the offer below which all responders reject converged to about 0.4. In contrast, the average offer converged to about 0.5. Empirical observations show a rather wide distribution of offers, with a considerable fraction of people indeed making the equipartition offer of 0.5.

The solution outlined above can be understood as follows. As the Golden Ratio is that number that is hardest to approximate by fractions of small integers, such an offer is hardest to evaluate by the responder in view of whether it is still near equipartition. Loosely speaking, she will not easily realize that she is disadvantaged.

The above explanation is not a mathematical proof in the strict sense. It is based on several assumptions. Two very crucial assumptions are that an offer of 0.5 would always be accepted while an offer of 1/3 would usually not. While the former assumption is very plausible and is used by many authors in the field, it requires scrutiny. In an iterative version of the Ultimatum Game, the responder would have considerable power. She could push up offers by rejecting first. It is likely that, in the iterative version, many equilibrium solutions exist. If, say $1-x =$

57 % have been offered repeatedly (that is $x < 0.5$!), the proposer can hardly lower offers in later rounds because the responder might reject until 57 % are offered again. Interestingly, in the study by Henrich et al. (2005), one ethnical group (the Lamalera in Indonesia) were found to make average offers larger than 0.5, notably 0.57. My experience is that some children also make offers slightly higher than 0.5 in order to make absolutely sure that it will be accepted, although children appear to be more modest than adults when being on the responder side (Proctor et al., 2013).

However, in this paper, we deal with the one-shot version of the game. Given that the people might assume that it is repeated, the solution will be influenced by learned rules of behaviour that exclude a behaviour pushing up offers unduly. Therefore, the proposer would not usually assume the responder to be that hard-boiled.

As the proposer needs to express her offer in integer or perhaps rational numbers rather than by an intuitively felt ratio, it is unlikely that she would propose 38,2... %. Rather, she would round to 40 % or 1/3. These options for $1-x$ correspond, in fact, to two Fibonacci approximations for the GR, notably 3/5 and 2/3. And so does the equipartition solution ½. This is reminiscent of the various approximations of the GR realized in phyllotaxis (Jean, 1994). Alternate leaves correspond to a ratio of ½, but other ratios of Fibonacci numbers can be observed on plants as well.

The question arises whether the GR solution is found in the mind of the proposer immediately by intuition or by iteration in correspondence to Fibonacci numbers. This is hard to answer and deserves further study. As mentioned above, the aesthetic appeal of the GR is realized by people by intuition rather than iteration. In the Ultimatum Game, though, we assume that some iteration is occurring because the cases (i) and (ii) mentioned above Eq. (5) are likely to be thought about by the proposer. In the case of iteration, it is unlikely that the proposer would think about the convergence and, thereafter, would go back a step and approximate the irrational number by a rational number, say 3/5. Rather, she would stop the iteration at 3/5.

The Ultimatum Game might even be relevant in microbiology. Several micro-organisms show the so-called sequential cross-feeding (Rozen and Lenski, 2000; Pfeiffer and Bonhoeffer, 2004). That is, a product excreted by one of them is taken up as a nutrient by the other. For example, some strain of the bacterium, *Escherichia coli* takes up glucose and converts it into acetate. This is taken up by another strain of the same species, which converts it further into $CO_2$ and water. During evolution, the two strains or species must come to an agreement about the metabolic intermediate to be excreted and taken up. It might have been ethanol (as is the case for other pairs of species) rather than acetate. If the responder rejects the offered intermediate, the proposer will not survive either or at least be harmed because substances such as acetate and ethanol are toxic in higher concentrations. As these intermediates are much poorer substrates than glucose, the observed offer is less valuable than 50 %.

Page and Nowak (2002) stress the importance of empathy in explaining the observations with humans. However, it need be clarified what is meant by empathy. It may mean anticipation of the strategy of the responder by thinking how would one decide when in her position. This is a usual assumption in analysing sequential games, although predicting the counterpart's moves is often quite difficult, especially in games with many consecutive steps like chess. Anticipation usually stops after thinking about two or three consecutive steps, which might explain an approximate solution (see above). If empathy means compassion (which would prevent unfair offers), it is not always justified because usually players are assumed to maximize their payoff. For example, compassion would prevent the defect strategy in the Prisoner's Dilemma, while exactly that strategy is often observed.

As most people assume that the game will be iterated, there is actually more than one reason why the proposer should not offer too little. First, she must be afraid of rejection and second, she might assume that in a future round of the game, she will be a responder and the proposer who was a responder before might retaliate.

The Golden Ratio is an irrational number. However, this need not imply that making an offer corresponding to it would be an irrational decision. The word irrational obviously has two different meanings. Interestingly, the mathematical meaning in defining a type of numbers has a pejorative connotation that may be questioned. Some "irrational" numbers may be understood in an easy and natural way. For example, the length of the diagonal of a right, equilateral triangle is intuitively comprehensible, although $\sqrt{2}$ is an "irrational" number. The same holds for the Golden Ratio.

**Acknowledgments**

I am very grateful to Karen Page (London), who was the first to play the Ultimatum Game with me (I as the proposer offered 40 %). I also thank Sebastian Germerodt and Günter Theißen (both Jena) for valuable feedback on the manuscript and Robert Schuster (Berlin) and Ramzi Suleiman (Haifa) for stimulating discussions.

**References**

K. Bohl, S. Hummert, S. Werner, D. Basanta, A. Deutsch, S. Schuster, G. Theißen, A. Schröter: Evolutionary game theory: molecules as players. *Mol. Biosystems* 10 (2014) 3066–3074.

C. Camerer, R.H. Thaler: Ultimatums, dictators and manners. *J. Econ. Persp.* 9 (1995) 209–219.

A. Cuyt, V.B. Petersen, B. Verdonk, H. Waadeland, W. B. Jones: Handbook of Continued Fractions for Special Functions. Springer, Berlin/New York 2008.

H. Gintis: Game Theory Evolving. Princeton University Press, Princeton 2000.

W. Güth, R. Schmittberger, B. Schwarze: An experimental analysis of ultimatum bargaining. *J. Econ. Behav. Organiz.* 3 (1982) 367-388.

J. Havil: The Irrationals. Princeton University Press, Princeton 2012.

J. Henrich, R. Boyd, S. Bowles et al.: "Economic man" in cross-cultural perspective: Behavioral experiments in 15 small-scale societies. *Behav. Brain Sci.* 28 (2005) 795–855.

J. Henrich, J.B. Silk: Interpretative problems with chimpanzee ultimatum game. *Proc. Natl. Acad. Sci. U.S.A.* 110 (2013) E3049.

J. Hofbauer, K. Sigmund: Evolutionary Games and Population Dynamics, Cambridge University Press, Cambridge 1998.

J. Iranzo, J. Román, A. Sánchez: The spatial Ultimatum game revisited. *J. Theor. Biol.* 278 (2011) 1-10.

R.V. Jean, Phyllotaxis. Cambridge University Press, Cambridge 1994.

K. Jensen, J. Call, M. Tomasello: Chimpanzees are rational maximizers in an ultimatum game. *Science* 318 (2007) 107-109.

D.A. King: The adaptive significance of tree height. *Am. Nat.* 135 (1990) 809–828.

T. Koshy: Fibonacci and Lucas Numbers with Applications, Wiley, New York 2001.

J. Maynard Smith: Evolution and the Theory of Games, Cambridge University Press, Cambridge 1982.

M.A. Nowak, K.M. Page, K. Sigmund: Fairness versus reason in the ultimatum game. *Science* 289 (2000) 1773-1775.

C.D. Olds, A. Rockett, P. Szusze: Continued Fractions. Mathematical Assn. of America, Washington., D.C. 1992.

K.M. Page, M.A. Nowak: A generalized adaptive dynamics framework can describe the evolutionary Ultimatum Game. *J. Theor. Biol.* 209 (2000) 173-179.

K.M. Page, M.A. Nowak: Empathy leads to fairness. *Bull Math Biol.* 64 (2002) 1101-1116.

T. Pfeiffer, S. Bonhoeffer: Evolution of cross-feeding in microbial populations. *Am. Nat.* 163 (2004) E126-E135.

D. Proctor, R.A. Williamson, F.B. de Waal, S.F. Brosnan: Chimpanzees play the ultimatum game. *Proc. Natl. Acad. Sci. U.S.A.* 110 (2013) 2070-2075.

A.E. Roth; V. Prasnikar; M. Okuno-Fujiwara; S. Zamir: Bargaining and market behavior in Jerusalem, Ljubljana, Pittsburgh, and Tokyo: An experimental study. *Amer. Econ. Rev.* 81 (1991) 1068-1095.

D.E. Rozen, R. E. Lenski: Long-term experimental evolution in *Escherichia coli*. VIII. Dynamics of a balanced polymorphism. *Am. Nat.* 155 (2000) 24–35.

R. Suleiman: Expectations and fairness in a modified ultimatum game. *J. Econ. Psychol.* 17 (1996) 531-554.

R. Suleiman: An aspirations-homeostasis theory of interactive decisions. (2014) http://vixra.org/abs/1403.0029